\documentstyle[aps,epsfig,preprint]{revtex} % one column
\tightenlines
\newcommand{\be}{\begin{equation}}
\newcommand{\ee}{\end{equation}}
\newcommand{\bea}{\begin{eqnarray}}
\newcommand{\eea}{\end{eqnarray}}

\newcommand{\noi}{\noindent}

\newcommand{\pst}{\protect\textstyle\scriptscriptstyle}

\newcommand{\bmr}{\mbox{\boldmath $r$}}

\newcommand{\wk}{\omega_k}

\newcommand{\bnu}{\mbox{\boldmath $\nu$}}
\newcommand{\bJ}{\mbox{\boldmath $J$}}
\newcommand{\bF}{\mbox{\boldmath $F$}}

\begin{document}
%%%%%%%%%%%%%%%%%%%%%%%%%%%%%%%%%%%%%%%%%%%%%%%%%%%%%%%%%%%%%%%%%%%%%%%%%%%

%%%%%%%%%%%%%%%%%%%%%%%%%%%%%%%%%%%%%%%%%%%%%%%%%%%%%%%%%%%%%%%%%%%%%%%%%%%
% TITLEPAGE
%%%%%%%%%%%%%%%%%%%%%%%%%%%%%%%%%%%%%%%%%%%%%%%%%%%%%%%%%%%%%%%%%%%%%%%%%%%

\preprint{UCY-PHY-99/3; DFTUZ/99/16}

\title{\bf Thermodynamics of self-gravitating systems with softened 
potentials}

\author{Eduardo Follana}
\address{Department of Natural Sciences, University of Cyprus, 
P.O. Box 20537, Nicosia CY-1678, Cyprus
\\ eduardo@dirac.ns.ucy.ac.cy}
\author{Victor Laliena}
\address{Departamento de F\'{\i}sica Te\'orica, Universidad de Zaragoza,
E-50009 Zaragoza, Spain\\
laliena@posta.unizar.es}
\date{February 22, 2000} 
%\pacs{PACS numbers: 05.20.-y, 05.90.+m, 05.70.Fh}

\maketitle

\begin{abstract}
The microcanonical statistical mechanics of a set of self-gravitating
particles is analyzed in a mean-field approach. In order to deal
with an upper bounded entropy functional, a softened gravitational
potential is used. The softening is achieved by truncating to N terms
an expansion of the Newtonian potential in spherical Bessel
functions. The order $N$ is related to the softening at short
distances.  This regularization has the remarkable property that it
allows for an exact solution of the mean field equation. It is found
that for N not too large the absolute maximum of the entropy coincides
to high accuracy with the solution of the Lane-Emden equation, which
determines the mean field mass distribution for the Newtonian
potential for energies larger than $E_c\approx -0.335 G M^2/R$. Below
this energy a collapsing phase transition, with
negative specific heat, takes place. The dependence of this result on the
regularizing parameter $N$ is discussed.
\end{abstract}

%\pacs{PACS numbers: 05.20.-y, 05.90.+m, 05.70.Fh}

%%%%%%%%%%%%%%%%%%%%%%%%%%%%%%%%%%%%%%%%%%%%%%%%%%%%%%%%%%%%%%%%%%%%%%%%%%%
% MAIN PART OF PAPER
%%%%%%%%%%%%%%%%%%%%%%%%%%%%%%%%%%%%%%%%%%%%%%%%%%%%%%%%%%%%%%%%%%%%%%%%%%%
%\narrowtext

%------------------------------------------------------------------------
\section{Introduction} 
%------------------------------------------------------------------------

The statistical mechanics of self-gravitating systems is amazing. It
has been studied since long ago by Antonov \cite{antonov}, Lynden-Bell
and Wood \cite{lbw}, Thirring \cite{th},  and Kiessling \cite{kiess},
among others
\cite{pad}. One reason for the interesting and peculiar behavior of
these systems is that they are thermodynamically unstable. The usual
thermodynamical limit exists only for those systems which are
thermodynamically stable \cite{linden}. For a system of $N_p$
classical particles interacting via a two body potential $\phi(r)$, a
sufficient condition for thermodynamical stability states that there
must exist a positive constant $E_0$ such that, for each configuration
$\{\bmr_1,\ldots,\bmr_{N_p}\}$, the following inequality is
obeyed~\cite{linden}:

\be 
\Phi(\bmr_1,\ldots,\bmr_{N_p})\;=\;\frac{1}{2}\sum_{i\neq j}
\phi(|\bmr_i-\bmr_j|)\;
\geq -N_p E_0\; . \label{stc}
\ee

\noi In contrast, self-gravitating systems do not possess a proper
thermodynamical limit. Moreover, due to the short distance singularity
of the gravitational potential, the entropy is not even well defined:
it diverges for any value of the energy \cite{antonov,pad}. To define
the thermodynamics of these systems the potential must be regularized
at short distances. This can be done in many different ways. Particles
endowed with a hard core is one possibility \cite{hcore1,hcore2}. 
In this case, the
potential is repulsive and singular at short distances. Other popular
choices are the so called softened potentials, which are smooth at the
origin. As shown by Thirring \cite{th}, the thermodynamical
instability is caused neither by the singularity nor by the long range
nature of the potential, but is due to the fact that the potential is
always attractive\footnote{In the case of hard core particles the
potential is repulsive at short distances and the thermodynamical
instability is actually due to the long range forces.}.  The essential
common feature of these purely attractive potentials is the appearance
of a phase transition separating a high energy homogeneous phase (HP)
from a low energy collapsing phase (CP) \cite{comp,posch,mila}.  The phase
transition takes place in an energy interval with negative
microcanonical specific heat. From the dynamical point of view, both
phases are also different: the single particle motion is
superdiffusive in the CP and ballistic in the HP \cite{latora,torcini}.
The dynamics and statistics of simple low dimensional models with long 
range attractive forces has been
studied in \cite{latora,torcini,ruffo,miller}. Their conclusions 
support the idea of a collapsing phase transition as in the Thirring model.
If angular momentum is conserved, the situation could be notably altered
\cite{victor}.

As mentioned, the usual thermodynamical limit does not exist for unstable
systems. To have well defined thermodynamics when the number of particles 
$N_p$ is huge, the following scaling must be considered: the potential
energy is rescaled by $1/N_p$, and then the energy and entropy scale with
$N_p$. It has been proved for the canonical ensemble that this scaling
reproduces mean field theory exactly in the limit $N_p\rightarrow\infty$
\cite{meanfield}. This means that correlations among two or more particles
vanish, and therefore the equilibrium state is characterized by a one particle
density only, which minimizes the free energy functional. Although we are
not aware of any rigorous proof, we shall assume here
that the same holds for the microcanonical ensemble, changing minimization
of the free energy by maximization of the entropy functional.

If the troubles caused by the short distance singularity are ignored,
it is possible to write down a mean field entropy functional for 
self-gravitating systems, which depends only 
on the particle density. This functional is not upper
bounded, and, therefore, has no absolute maximum, a reflection of the fact
that the entropy is not defined in the finite system. For energies larger
than $E_c \approx -0.335 GM^2/R$ there is however a local maximum. Below
this energy, no local maximum of the entropy exists
\cite{antonov}. This fact was explained in terms of a transition from
the homogeneous isothermal sphere behavior to the CP at $E_c$. The
transition produces negative specific heat, and was called the
gravo-thermal catastrophe \cite{lbw}. Very recently, it has been
pointed out that the low energy phase might be described by a
spherically non-symmetric deformation of the singular solution of the
isothermal Lane-Emden equation \cite{nieu}. 

The gravitational potential must be modified at short distances to
make equilibrium statistical mechanics applicable to self-gravitating
systems. As shown by Kiessling for the canonical ensemble
\cite{kiess}, in the limit where the classical gravitational potential
is recovered the equilibrium state approaches a particle distribution
with all particles collapsing at a single point. The behavior of the
system will depend on the scales at which the regularization is
effective. There might be regularized potentials which, in the mean
field limit, produce the global maximum of their associated entropy
close to the solution of the isothermal Lane-Emden equation for
energies $E\geq E_c$. If this is the case, a collapsing transition
should occur at some energy close to $E_c$. The CP is expected to be
very sensitive to the details of the regularization at short distances
and the HP almost insensitive to it. On the other hand, if the
regularization is effective only at very short distances, the
solutions of the isothermal Lane-Emden equation will be global maxima
of the entropy only at very high energies, and therefore the
collapsing transition will take place at some energy much larger than
$E_c$. The critical energy will be higher the smaller the scale at
which the regularized potential differs significantly from the
unregularized one. A completely similar picture was rigorously
established by Kiessling for the canonical ensemble \cite{kiess}.

In this paper we introduce a convenient new softening procedure for
the regularization of the gravitational potential, and we investigate
the consequences in the microcanonical thermodynamics of self-gravitating 
systems. The rest of the article is organized as follows:
in Sec. II we introduce the family of potentials to be studied; in
Sec. III we derive the mean field equation and its general solution in
terms of a set of algebraic equations; Sec. IV is devoted to the
discussion of the results and Sec. V to summarize the conclusions.

%------------------------------------------------------------------------
\section{Softened potential \label{sec:pot}} 
%------------------------------------------------------------------------

As mentioned in the introduction, the short distance singularity of
the gravitational potential causes many troubles. What is more, for a
real system such singularity is not physical, since at short distances
new physics must be taken into account. Thus, the potential should be
modified at short distances to avoid the singularity. 
In simulations of cosmological
problems a widely used potential is the so called Plummer softened
\cite{plummer,bintre}:

\be
\phi (r) = -\frac{G M^2}{\sqrt{r^2 + \sigma^2}} \, . \label{plummer}
\ee

\noi For $r\gg\sigma$, (\ref{plummer}) coincides with the
gravitational potential. Other softened potentials are those known as
spline softened \cite{hern}.  The equilibrium thermodynamics of
systems with these softened potentials has been studied in \cite{spsoft},
and the dynamical effects of softening were considered in \cite{romeo}.
The form of the potential at short distances is arbitrary to a large extent, 
since we do not know how the new interactions modify it.

The problems with the singularity of the gravitational potential in
statistical mechanics disappear if the equilibrium distribution is
modified appropriately. For instance, if the approach to equilibrium is
collisionless, via violent relaxation, the equilibrium state is
described by the Lynden-Bell statistics \cite{lbstat}, whose
one-particle distribution is of Fermi-Dirac type, and produces an
effective repulsion at short distances \cite{chavanis}.  

The regularized potential we propose, which, as we shall discuss, has
several remarkable features making it very convenient for
thermodynamical purposes, is based on the following 
identity\footnote{Equation (\ref{expansion}) follows immediately from
the sine series expansion of the constant function, $f(x)=1$, in the
interval $(0,1)$.}:

\be
\frac{1}{x} \;=\; 4\sum_{k=1}^{\infty}\,\frac{\sin[(2k-1)\pi x]}
{(2k-1)\pi x}\, ,\hspace{1.5 truecm} 0\,<\, x\,<\,1\: .
\label{expansion}
\ee

\noi The singularity at the origin is removed by truncating the series
to a given order $N$. Let us consider a system of $N_p$ particles
confined within a sphere of radius $R$ in 3 dimensions. The maximum
distance between two particles is $2R$. Hence, our potential must
represent $1/r$ for distances $0\ll r<2R$. Therefore, we choose the
following interaction energy between two particles of mass $m$ located
at $\bmr$ and $\bmr^\prime$:

\be
\phi(|\bmr-\bmr^\prime|)\;=\;-\frac{G m^2}{R}\,2\,\sum_{k=1}^N\,
\phi_k(|\bmr - \bmr^\prime|/R) \, ,
%\frac{\sin(\wk |\bmr - \bmr^\prime|/R)}{\wk |\bmr - \bmr^\prime|/R}
\label{pot}
\ee

\begin{figure}[htb]
\begin{center}
\epsfig{file=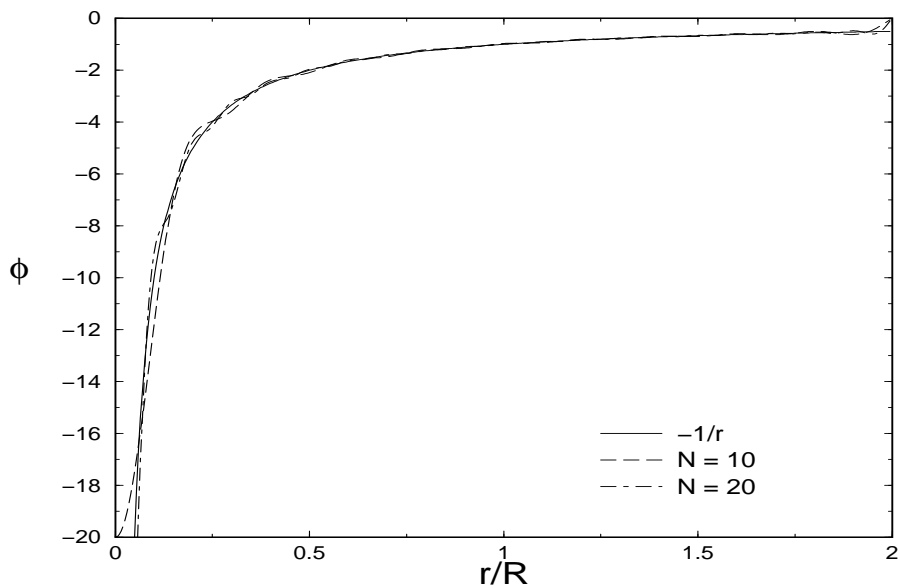,height=12cm,width=8cm,angle=-90}
\end{center}
\caption{Newtonian and regularized potentials with
 $N=10$ and $20$, in units of $GM^2/R$.}
\label{fig:pot}
\end{figure}

\noi where $\phi_k(x)=\sin(\wk x)/(\wk x)$ are spherical Bessel
functions of order zero and $\wk=(2k-1)\pi/2$.  Fig. \ref{fig:pot}
displays the singular potential and the regularized potential with
$N=10$ and $N=20$. A similar expansion has been used to introduce
simple models in low dimensions which allow to perform numerical
simulations of systems with long range attractive forces with CPU time
growing only as the number of particles \cite{latora,torcini,ruffo}.
   
What is remarkable of (\ref{pot}) is that each term obeys the
following differential equation:

\be
(\,\nabla^2 \:+\: \frac{\wk^2}{R^2}\,)\,\phi_k(r/R) \;=\; 0 \, ,
\ee

\noi
so that the potential (\ref{pot}) verifies

\be
{\cal D}_N\,\phi(r) \;=\; 0\, , 
\label{depot}
\ee

\noi 
where

\be
{\cal D}_N\;=\;\prod_{k=1}^N\,(\,\nabla^2 \:+\: \frac{\wk^2}{R^2}\,) \, . 
\ee

\noi
This relation will prove very useful in the mean field analysis of the
next section.

%------------------------------------------------------------------------
\section{Mean Field analysis \label{sec:mfa}} 
%------------------------------------------------------------------------

It is well known that long range forces suppress fluctuations, and
thus in these cases a mean field analysis is accurate or even
exact. We expect that, dealing with an unstable system in the scaling
regime described in the introduction, the description of the
thermodynamical state in terms of a one particle density, neglecting
two or more particle correlations, gives the essential physical
behavior \cite{meanfield}. We will derive in this section the form of
the mean field equation and of the corresponding thermodynamic
quantities for a system whose dynamics is governed by a potential of
the form (\ref{pot}).

\subsection{Mean Field Equation}

Let us consider a system of particles enclosed in a spherical region 
of radius $R$ and volume $V=4/3\pi R^3$, with a total mass $M$
distributed according to a smooth density $\rho(\bmr)$, normalized
such that $\int d^3r\rho(\bmr)=1$, and interacting via a two body
central potential \mbox{$\phi(|\bmr-\bmr'|)$}. If the potential is smooth,
the entropy per particle in the microcanonical ensemble can be written
in terms of the particle density as \cite{pad}:

\be
{\cal S}\; =\; -\,\int\,d^3r\,\rho(\bmr)\,\left[\,\ln V\rho(\bmr)\,-\,1\,
\right]\:+\frac{3}{2}\,\ln\,\left(\,E\,-\,\Phi\,\right)\; , 
\label{mfs}
\ee

\noi where $E$ is the total energy and $\Phi$ is the potential energy:

\be
\Phi[\rho]\; =\; \frac{1}{2}\,
\int\,d^3r\,d^3r'\,\rho(\bmr)\,\phi(|\bmr-\bmr'|)\,\rho(\bmr')\, . \label{epot}
\ee

\noi The volume $V$ entering the first term of the r.h.s in
(\ref{mfs}) has been included to make the entropy look dimensionally
correct, and plays no significant role since it only shifts the
entropy by a constant.  The physical density is the absolute maximum
of (\ref{mfs}) under the constraint $\int \rho = 1$. Differentiating
with respect to $\rho$ we arrive at the following integral equation:

\be
\ln V\rho(\bmr)\;=\;\mu\:-\:\frac{3}{2}\beta\int d^3r'\,\phi(|\bmr-\bmr'|)
\rho(\bmr')\, ,
\label{mfe}
\ee

\noi where $\beta=1/(E-\Phi)$ and $\mu$ is the Lagrange multiplier for
the constraint $\int\rho=1$. Defining $\nu(\bmr)$ by

\be
\rho(\bmr)\;=\;\frac{1}{V}\,\exp[\mu + \nu(\bmr)] \, , \label{nudef}
\ee

\noi
the constraint is solved by taking

\be
e^\mu = \frac{V}{\int d^3r e^{\nu(\bmr)}} \, . \label{musol}
\ee

\noi
Substituting (\ref{nudef}) and (\ref{musol}) in (\ref{mfe}), we obtain 
for $\nu(\bmr)$:

\be
\nu(\bmr)\;=\;-\frac{3}{2}\,\beta\:\frac{\int d^3r'\,\phi(|\bmr-\bmr'|)\,
e^{\nu(\bmr')}}{\int d^3r'\,e^{\nu(\bmr')}} \, . \label{ienu}
\ee

If we take for $\phi(r)$ the Newtonian potential, we know that the
entropy is not well defined. Nevertheless, it is still possible to
start formally with the entropy functional (\ref{mfs}), which gives a
finite result for any smooth distribution $\rho(\bmr)$, but is
unbounded (see \ref{subs:ndep}). There can still exist local maxima,
which are then solutions of (\ref{ienu}). By expanding the right-hand
side of this equation in a series of spherical Bessel functions and
truncating after N terms, one would obtain results equivalent to the
ones we get using the softened potential.

If we now particularize (\ref{ienu}) to the softened potential
(\ref{pot}), we see that $\nu(\bmr)$ obeys the same differential
equation (\ref{depot}) as the potential. Imposing rotational symmetry
on $\nu$ \cite{antonov,lbw}, we obtain the following ordinary
differential equation:

\be
\prod_{k=1}^N\,\left(\frac{d^2}{dr^2}\,+\,\frac{2}{r}\frac{d}{dr}\,+\,
\frac{\wk^2}{R^2}\right)
\:\nu(r)\;=\;0 \, .
\ee

\noi The general solution of this equation is a linear combination of
$\{\sin(\wk r)/r\}$ and $\{\cos(\wk r)/r\}$. The cosines should be
absent from the solution since (\ref{ienu}) implies that $\nu$ is
smooth. (Only at $T=0$, i.e., $\beta=\infty$, is $\nu$ singular.)
Indeed, it is shown explicitly in appendix A that the 
solution of (\ref{ienu}) can be written as

\be
\nu(r)\;=\;\sum_{k=1}^N\,\nu_k\,\phi_k(r/R) \, ,  \label{sol}
\ee

\noi
where $\nu_k$ are $N$ numerical coefficients determined by the following set
of equations:

\be
\nu_k \;=\; 3\,\beta\,\frac{GM^2}{R}\:\frac{\int_0^1\,dx\,x^2\,\phi_k(x)\,
\exp\{\sum_k\nu_k\phi_k(x)\}}
{\int_0^1\,dx\,x^2\,\exp\{\sum_k\nu_k\phi_k(x)\}} \,\, . \label{eqnu}
\ee

\noi The integral equation (\ref{ienu}) has been reduced to a system
of $N$ non-linear algebraic equations with $N$ unknowns. It can be
solved by iteration, for instance with a Newton algorithm (see appendix
B for a summary of the method used in this work).

\subsection{Thermodynamical quantities}

Using formula (\ref{apfor}) of appendix A, it is straightforward to
verify that, for a spherically symmetric mass distribution
$\exp(\mu+\nu(r))$, the potential energy is given by:

\be
\Phi \;=\; -\frac{GM^2}{R}\,
\sum_{k=1}^N\,\left[\,\frac{\int_0^R\,dr\,r^2\,\phi_k(r/R)\,
e^{\nu(r)}}{\int_0^R\,dr\,r^2\,e^{\nu(r)}}\,\right]^2\, .
\ee

\noi
For an equilibrium distribution of the form (\ref{sol}) , Eq. (\ref{eqnu})
implies

\be
\Phi \;=\; -\frac{1}{9\beta^2}\,\frac{R}{GM^2}\,\sum_{k=1}^N\,\nu_k^2 \, .
\ee

\noi
Hence, the total energy is

\be
E\;=\;\frac{1}{\beta}
\:-\:\frac{1}{9\beta^2}\,\frac{R}{GM^2}\,\sum_k\nu_k^2 \, .
\ee

From (\ref{mfs}) we easily obtain the equilibrium entropy:

\be
{\cal S} \;=\;-\,\ln\left(\int_0^1\,dx\,x^2\,
\exp(\sum_k\,\nu_k\phi_k(x))\right)\:+\:
\frac{R}{GM^2}\,\frac{\sum_k\nu_k^2}{3\beta} \:-\:\frac{3}{2}\ln\beta \, .
\ee

\noi
Since the entropy is stationary under variations of the mass distribution,
the inverse temperature is $1/T\equiv\partial {\cal S}/\partial E
=\beta = 1/(E-\Phi)$.

%------------------------------------------------------------------------
\section{Results \label{sec:res}} 
%------------------------------------------------------------------------

In order to present specific numerical results, it is convenient to
work with dimensionless quantities. We measure the energy in units of
the characteristic energy, $G M^2/R$, where $M$ is the total mass and
$R$ the radius of the confining sphere. The dimensionless energy is
then $\epsilon=ER/(GM^2)$. Any other quantity with dimensions of
energy (like the temperature $1/\beta$ and the potential energy
$\Phi$) must be also understood to be expressed in units of $G M^2/R$,
and, similarly, magnitudes with dimensions of length are given in
units of $R$. As a matter of terminology, we shall call Newtonian
potential (NP) to the unregularized potential, $-GM^2/r$, Newtonian
entropy (NE) to its corresponding entropy, regularized potential (RP)
to the potential regularized by Eq. (\ref{pot}), and regularized
entropy (RE) to its associated entropy.

\subsection{ \ N = 10  \label{subs:n10}}

\begin{figure}[htb]
\begin{center}
\epsfig{file=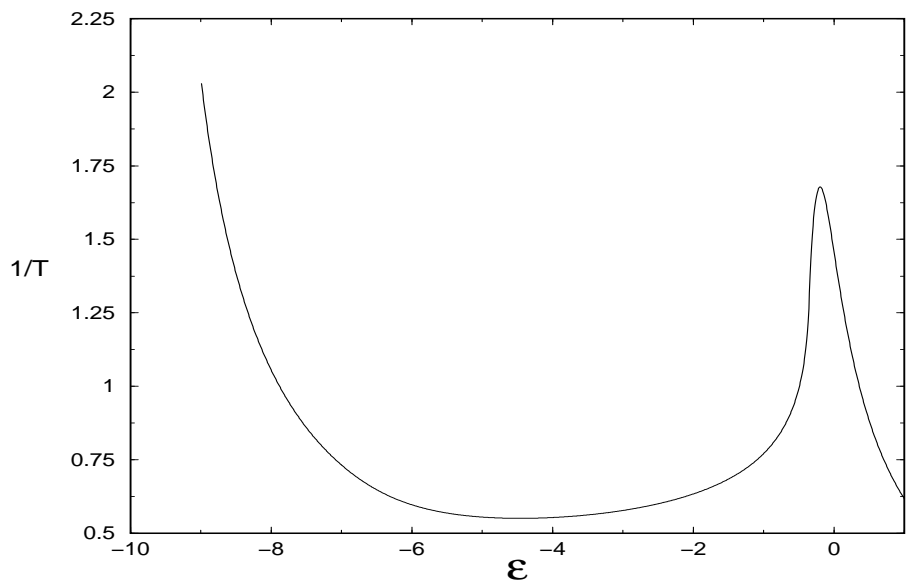,height=12cm,width=8cm,angle=-90}
\end{center}
\caption{Inverse temperature $1/T$ versus energy $\epsilon$ for 
$N = 10$.}
\label{fig:cal}
\end{figure}

For values of $N$ not too large, computations are easy. Let us
describe the case $N=10$ in detail. (See appendix B for
a summary of the numerical methods used in this work).
The solution of Eq. (\ref{eqnu})
as a function of the energy $\epsilon$ provides all thermodynamic
functions.  For each value of $\epsilon$ we found only one solution,
which should then be the absolute maximum of (\ref{mfs}). We shall
return to this point later on, in
Sec. \ref{subs:ndep}. Fig. \ref{fig:cal} displays the inverse
temperature $\beta=1/T$ versus $\epsilon$. In the thermodynamics of
stable systems, this function must be monotonically decreasing, since
the entropy is a convex function of the energy \cite{vh}. In the present 
case, however, $\beta$ decreases with $\epsilon$ in the low and high
energy regimes, but it increases for $\epsilon$ in
$(-4.46,-0.2)$ and, consequently, the specific heat is negative in
this energy interval. This is a consequence of the instability of the
system. 

As is usually the case with these systems \cite{th}, the negative
specific heat region is associated with a transition to a collapsed
phase. To investigate this, let us define an order parameter
$\kappa=R_0/R$, where $R_0$ is the radius of the sphere centered at
the origin which contains 95\% of the mass (of course the value of
95\% is arbitrary).  Fig. \ref{fig:rad} displays $\kappa$ versus
$\epsilon$. At $\epsilon=\infty$ the mass is distributed
homogeneously, and then $\kappa=(0.95)^{1/3}\approx 0.9830$. When the
energy is reduced, $\kappa$ decreases monotonically and slowly. Notice
the anomaly at $\epsilon\sim-0.335$; we shall discuss it in
Sec. \ref{subs:le}.  The collapsing order parameter $\kappa$ varies
abruptly in the region where the specific heat is negative. It decays
from $\kappa\sim 0.95$, corresponding to an homogeneous phase, to
$\kappa\sim 0.1$. In the later case, the mass distribution consists of
a small dense core and an homogeneous tenuous halo. 

The results of this section are similar to those found by regularizing
the potential with hard core repulsions \cite{hcore1,hcore2}, and to those 
derived from the
Lynden-Bell statistics applied to the unregularized potential 
\cite{chavanis}. It is remarkable that different regularizations lead
to similar results.
 
\subsection{ \ Newtonian potential \label{subs:le}}

It is interesting to compare the maximum of the RE with the local
maximum of the NE.  Substituting
$\phi(|\bmr-\bmr'|)=-GM^2/|\bmr-\bmr'|$ in (\ref{ienu}), and using
the fact that this potential is a Green function of the Laplacian, we
get the following differential equation

\be
\nabla^2 \nu(\bmr) \;=\; -\frac{3}{2}\,\beta\,e^\mu\,\exp(\nu(\bmr)) \, ,
\ee

\noi which, for spherically symmetric $\nu$, is equivalent to the
isothermal Lane-Emden equation \cite{emden,chan,lbw}:

\be
\frac{d^2\nu(r)}{dr^2}\,+\,\frac{2}{r}\,\frac{d\nu(r)}{dr}\,+\,
\frac{3}{2}\beta e^\mu\,e^{\nu(r)}\;=\; 0 \label{lee}
\ee

\noi The proper solutions of Eq. (\ref{lee}), with $\beta$ and $\mu$
such that $\beta=1/(\epsilon-\Phi)$ and $\mu=-\ln\int dr r^2
\exp\nu(r)$, give local maxima of the entropy if $\epsilon>-0.335$
\cite{antonov,lbw}. The high energy phase should depend only weakly on
the form of the potential at short distances. Therefore, the maximum
of the RE in the high energy phase might be an approximation to the
local maximum of the NE given by Eq. (\ref{lee}).  This is indeed the
case.

To see how close the maximum of the RE is to the appropriate solution of
eq. (\ref{lee}), we define a distance between functions by
$D=\max\{|\nu_{\pst N=10}(r) - \nu_{\pst LE}(r)|\}$, where the
subscripts indicate the solutions of Eq. (\ref{eqnu}) with $N=10$, and
of Eq. (\ref{lee}) respectively.  For $\epsilon> -0.335$, i.e., when
the Lane-Emden Eq. determines a local maximum of the NE,
$D<10^{-4}$. The absolute maximum of the RE is indeed a very good
approximation to the local maximum of the NE.

\begin{figure}[htb]
\begin{center}
\epsfig{file=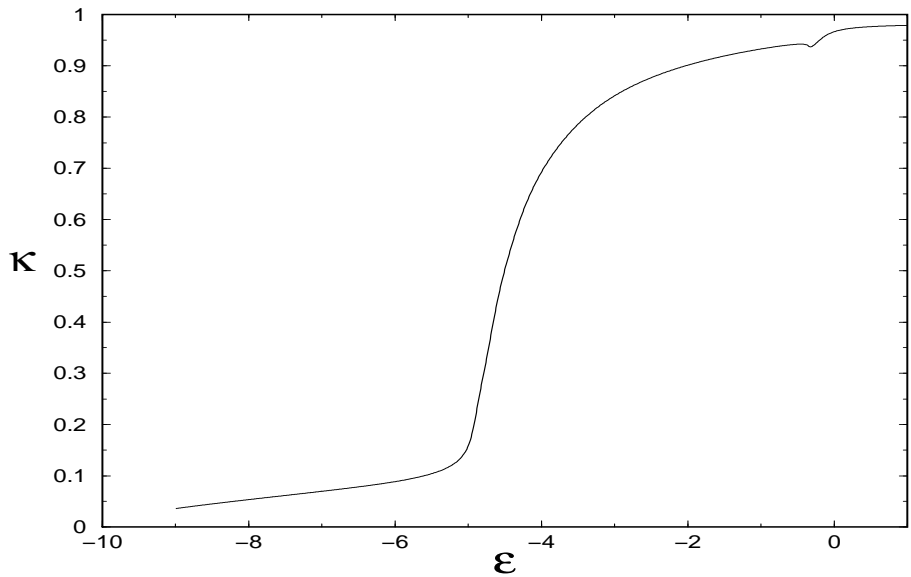,height=12cm,width=8cm,angle=-90}
\end{center}
\caption{Order parameter of collapsing (see text, Sec. \ref{subs:n10})  
versus energy $\epsilon$ for $N = 10$.}
\label{fig:rad}
\end{figure}

Now, we can understand the anomaly in $\kappa$ around
$\epsilon\sim-0.335$, which was mentioned in Sec. \ref{subs:n10} and
which can be appreciated in Fig. \ref{fig:rad}. At this point, which
is close to the energy at which the solutions of the Lane-Emden
eq. cease to be local maxima of the NE, the nature of the maximum of
the RE also changes, originating anomalies such as the pick in $1/T$
(Fig. \ref{fig:cal}) and the fissure in $\kappa$ (Fig. \ref{fig:rad}).

The effect of the regularization is to deform the entropy functional
dramatically for mass distributions $\rho(r)$ which are very
concentrated at the origin.  These distributions get a huge amount of
negative entropy after softening the potential, at least for $N=10$,
in such a way that there are no maxima of the RE close to them.  On
the other hand, the entropy of smooth distributions which are not
concentrated is sensitive to the global form of the potential rather
than to the short distance details.  Therefore, these distributions
have similar NE and RE, and they essentially do not feel the
regularization. The solutions of the Lane-Emden Eq. belong to this
class, and, consequently, close to them there is a local maximum of
the RE which is indeed the global maximum for not too large $N$, in
particular for $N=10$.

\subsection{ \ $N$ dependence \label{subs:ndep}}

The NP can be arbitrarily well approximated at short distances by a RP
with $N$ sufficiently large. Consequently, the maximum of the RE close 
to the local maximum of the NE will attain the later in the
$N\rightarrow\infty$ limit, and, obviously, must cease to be the global
maximum of the RE and become a local one for some value of $N$, which
will be denoted by $N_c$. Since the maximum of the entropy depends on 
the energy,  $N_c$ is a function of $\epsilon$. In principle, we can
compute $N_c(\epsilon)$ by solving Eq. (\ref{eqnu}) for large values
of $N$. In practice, however, this is very difficult and we must
content ourselves with an estimate of $N_c$.

To get the estimate, let us first analyze how matter distributions 
with arbitrarily high NE
can be built. There is an upper bound for the entropy functional
(\ref{mfs}) if the potential energy (\ref{epot}) is bounded from below
($\Phi\geq\Phi_{\pst min}$ for any $\rho(r)$):

\be
{\cal S}\; \leq \; 1\:+\:\frac{3}{2}\,
\ln(\epsilon-\Phi_{\pst min}) \, .
\label{bound}
\ee

\noi In the case of the RP, $\Phi_{\pst min}=-N$.  Since the entropy
has an upper bound, it is reasonable to assume that it has a global
maximum given by a regular function $\nu(r)$ of the form (\ref{sol}),
with coefficients $\nu_k$ verifying (\ref{eqnu}).  The potential
energy associated to the NP has no lower bound and therefore
$\Phi_{\pst min}=-\infty$. Hence, (\ref{bound}) does not provide
an upper bound for the NE. Indeed, it is straightforward to verify
that the distribution

\be
\rho(r)\;=\;  \left\{
\begin{array}{ll}
\frac{3\alpha}{4\pi r_0^3} & \hspace{1 cm} 0 \, < \, r \, < \, r_0 \\
\frac{3(1-\alpha)}{4\pi(1-r_0^3)} & \hspace{1 cm} r_0 \, < \, r \, < \, 1 \\ 
\end{array}
\right.  \label{divden}
\ee

\noi with $0<\alpha<1$, has arbitrarily large entropy when we take the
limit $r_0\rightarrow 0$, while maintaining $\alpha \ln r_0$
constant\footnote{Notice that in the limit $r_0\rightarrow 0$ with
$\alpha \ln r_0$ constant only an infinitesimal amount of matter
collapses, while the rest is homogeneously distributed.  This
reflects the fact that it is enough that two particles (hard
binaries) become arbitrarily close to make the potential energy
arbitrarily negative, and therefore the kinetic energy arbitrarily
large. However, as they constitute only two degrees of freedom, their
contribution to the purely configurational entropy term $\int\rho(\ln
V\rho-1)$ of (\ref{mfs}) is negligible.}. This is true for any value
of $\epsilon$.  We shall call these distributions, for any values of
$\alpha$ and $r_0$, special distributions (SD). If $N$ is large
enough, there are SD with larger RE than the maximum close to the 
solution of the Lane-Emden equation..

As already claimed, it is very difficult to get the solutions of eq.
(\ref{eqnu}) for large values of $N$. To overcome this problem and
obtain an estimate of $N_c$, we shall study the restriction of the RE
to particle distributions of the form (\ref{divden}) (SD). In such a
way, we have a RE which depends only on two parameters, $\alpha$ and
$r_0$. Now, the maximization of this entropy with respect to $\alpha$
and $r_0$ is an easy task, even for very large values of
$N$. Obviously, the smaller value of $N$ for which the maximum of the
restricted RE is larger than the RE of the corresponding solution of
the Lane-Emden equation will give an estimate of $N_c$. Strictly
speaking, this estimate is an upper bound on $N_c$.

\begin{figure}[htb]
\begin{center}
\epsfig{file=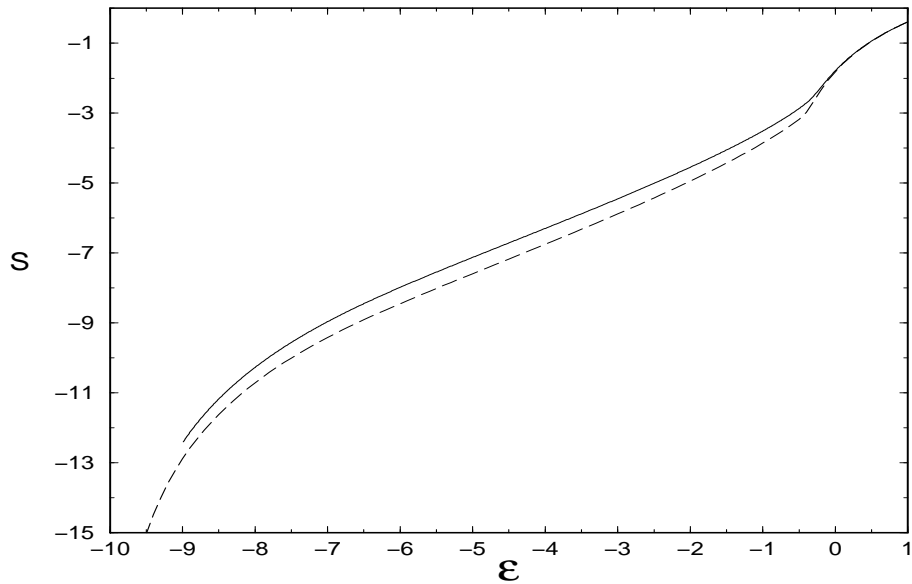,height=12cm,width=8cm,angle=-90}
\end{center}
\caption{Entropy $\cal{S}$ versus energy $\epsilon$ for $N = 10$. The solid 
line is the entropy of the solution of Eq. (\ref{eqnu}), and the dashed one
correspond to the maximum entropy of distributions of the 
form (\ref{divden}).}
\label{fig:ent}
\end{figure}

Before analyzing the behavior with $N$, let us look again to the 
$N=10$ case, for which computations are easy. Fig. \ref{fig:ent} 
displays the maximum of the restricted RE and the RE of the
corresponding solution of (\ref{eqnu}), both for $N=10$, as a 
function of $\epsilon$. The later distribution has always larger RE 
than any SD. This, besides the fact that we did not find other solutions
by varying the initial guess, confirms that for each $\epsilon$ only a
local maximum of the $N=10$ RE exists. It is, obviously, the global
maximum.

To investigate the behavior with $N$, we computed the estimate of 
the critical $N_c$ for several values of $\epsilon$. 
As it could have been anticipated, $N_c$ grows with $\epsilon$.  
Table \ref{tab} shows the
results. Column one displays $\epsilon$, column two the entropy of the
solutions of the Lane-Emden eq., column three the maximum of the 
restriction of the RE to SD for the estimated $N_c$, and column four 
the estimated $N_c$.  It is apparent that,
in the high energy phase ($\epsilon\geq -0.335$), we must go to $N$
larger than 30 to see global maxima different from the solutions of
the Lane-Emden equation. It is worth noting that a similar scenario
was rigorously established  by Kiessling for the equilibrium state of 
self-gravitating systems in contact with a thermal bath \cite{kiess}.

%------------------------------------------------------------------------
\section{Conclusions \label{sec:con}} 
%------------------------------------------------------------------------

To define the thermodynamics of gravitational systems properly, the
Newtonian potential must be regularized at short distances, removing
its singularity. Only then is the entropy well defined, or, in a mean
field approach, the entropy functional upper bounded. One way to
introduce a regularization is by softening, i.e., by making the
potential smooth at short distances while keeping it basically
unchanged at long distances. There are infinitely many ways to achieve
that. One interesting possibility is given by the truncation of the
expansion of the gravitational potential in spherical Bessel functions
to a given order $N$, as in Eq. (\ref{pot}). This regularization has
the virtue of reducing the mean field integral equation to a system of
$N$ algebraic equations with $N$ unknowns. This simplifies
considerably the solution of the problem.

The result which emerges from this approach is the following: if the
regularization is mild enough, $N<30$, the system undergoes a phase
transition separating a high energy homogeneous phase from a low
energy collapsed phase. In the high energy phase, the mass
distribution and the thermodynamic quantities are those of an
isothermal sphere.  Quantitatively, they are very close to the
solutions of the corresponding Lane-Emden equation. The low energy
phase is characterized by a mass distribution consisting of a dense
core surrounded by a tenuous halo. As usual in these cases \cite{th},
the transition from the HP to the CP takes place in an energy interval
with negative specific heat, an indication of the thermodynamical
instability of the system. These results are remarkably similar to
those found with a different regularization (hard core spheres)
\cite{hcore1,hcore2}, and with those derived from the Lynden-Bell statistics
applied to the unregularized potential \cite{chavanis}. We can then 
conclude that the thermodynamics is not very sensitive to the form
of the regularization.
 
The effect of a mild regularization is to deform the entropy functional
in such a way that, in the high energy phase, the global maximum of
the entropy with the regularized potential is very close to the local
maximum with the unregularized potential. The analysis based on the
Lane-Emden equation is therefore very accurate and the conclusions
extracted from it hold. If the potential is too sharp at
short distances, $N>30$, we expect also a collapsing transition, which
however will take place at a much higher energy than the one
predicted by the
analysis of the stability of the solutions of the Lane-Emden equation.
Below the critical energy the global maximum of the RE will not be
in the vicinity of the solution of the Lane-Emden equation, where, 
nevertheless, there will be a local maximum of the RE. Besides describing
such metastable states, the Lane-Emden equation might be physically 
relevant in diluted self-gravitating systems with an interparticle 
distance high enough to be insensitive to a truncation of the expansion
of the gravitational potential in spherical Bessel functions, Eq. 
(\ref{pot}), to 30 terms.

Finally, let us comment on the structure of the low energy microcanonical
equilibrium state when the regularized potential is very sharp 
($N>30$) at short 
distances. In this regime there are high entropy mass
distributions consisting of a small amount (infinitesimal when
$N\rightarrow\infty$) of matter condensed and the rest homogeneously
distributed. This might indicate that at these scales the system is
not well described by a smooth density, and granularity is playing a
major role.

\appendix

\section{\label{app:eq}}

Let us show that the solutions of Eq. (\ref{ienu}) for $\nu$
rotationally symmetric are of the form (\ref{sol}) if the potential is
$\phi(r)=\sum_{k=1}^N\phi_k(r)$. Introducing spherical coordinates,
eq. (\ref{ienu}) can be written

\be
\nu(r)\;=\;\frac{3\beta GM^2}{R}\,\sum_{k=1}^N\:
\frac{\int_0^1 dr' \,r'\, e^{\nu(r')} \int_0^\pi d\theta\,
\sin\theta\phi_k(\sqrt{r^2 + r'^2 - 2 r r' \cos\theta})}
{2\int_0^1 dr' r'^2  e^{\nu(r')}} \, . \label{apfor1}
\ee

\noi
The integral in $\theta$ can be readily performed, and gives

\be
\int_0^\pi d\theta\sin\theta
\frac{\sin[\wk (r^2 + r'^2 - 2 r r' \cos\theta)^{1/2}]}
{\wk(r^2 + r'^2 - 2 r r' \cos\theta)^{1/2}} \;=\;
2\phi_k(r)\phi_k(r')\, . \label{apfor}
\ee

\noi
Eqs. (\ref{apfor1}) and (\ref{apfor}) imply 
$\nu(r) \; = \; \sum_{k=1}^N\,\nu_k\,\phi_k(r)$,
with the coefficients $\nu_k$ determined by (\ref{eqnu}), QED.

\section{ \label{app:num} }

Since the numerical solution of Eq. (\ref{eqnu}) is central in this work,
we shall outline in this appendix the method used to solve it. The problem
is to find the roots of a vector function defined in a multidimensional
space. Eq. (\ref{eqnu}) can be written as

\be
F_i(\nu_1,\ldots,\nu_N)\;=\;0 \label{app:eq}
\ee

\noi
with $i=1,\ldots,N$. We shall use matrix notation and denote by
{\bnu } the complete set $(\nu_1,\ldots,\nu_N)$ and
by $\bF$ the vector $(F_1,\ldots,F_N)$.

To solve systems of equations like (\ref{app:eq}) we choose the
Newton-Raphson method, which works as follows \cite{numerical}: provided 
we have an
initial guess ${\bnu }$, which is close to the solution of 
(\ref{app:eq}),
we can expand $F_i$ in Taylor series in a neighbourhood of ${\bnu }$:

\be
F_i({\bnu } + \delta{\bnu })\;=\;F_i({\bnu })\:+\:
\sum_{j=1}^N\,J_{ij}\,\delta\nu_j
\:+\:O(\delta{\bnu}^2)
\ee

\noi
where $J_{ij}=\frac{\partial F_i}{\partial \nu_j}$ is the Jacobian matrix.
In matrix notation we have:

\be
{\bF}({\bnu } + \delta{\bnu })\;=\;{\bF}({\bnu })\:+\:
{\bJ}\cdot\delta{\bnu }\:+\:O(\delta{\bnu}^2)
\ee

\noi
By neglecting terms of order higher than linear in $\delta{\bnu }$ and by
setting ${\bF}({\bnu }+\delta{\bnu})=0$, we obtain a set of linear equations 
for the corrections $\delta{\bnu }$ that move each function $F_i$ closer to
zero simultaneously:

\be
{\bJ}\cdot\delta{\bnu }\;=\;-{\bF}
\ee

\noi
This linear equation is a standar problem in numerical linear algebra and
can be solved by LU decomposition. The corrections are then added to the 
initial guess,

\be
{\bnu}_{\pst new}\;=\;{\bnu }\:+\:\delta{\bnu}
\ee

\noi and the process is iterated to convergence. It is possible to
show that the method converges always provided the initial guess is
close enough to the root. It can also spectacularly fail to converge
indicating (though not proving) that the putative root does not exist
nearby. To avoid problems with the poor global convergence of the
method, we started with many different initial guess. We always found
convergence to the same solution, except when $N$ was larger than
$30$, where we found only convergence at high energy. We never got two
different solutions (whithin our convergence criterion, see below) by
starting at two different points.

Numerically, a convergence criterion is necessary. We stoped
computations when one of this two conditions

\be
\sum_{i=1}^N \left| \delta\nu_i \right|\:<\:10^{-10}
\hspace{0.75 cm} {\rm or } \hspace{ 0.75 cm}
\sum_{i=1}^N \left| F_i \right|\:<\:10^{-10}
\ee

\noi
was verified.

Each time the function $F_i$ was called, several integrals entering
eq. (\ref{eqnu}) were performed numerically, using a Romberg algorithm
\cite{numerical}. The integrands are smooth functions and it was
possible to achieve a high precision with a relatively modest
numerical effort.

%%%%%%%%%%%%%%%%%%%%%%%%%%%%%%%%%%%%%%%%%%%%%%%%%%%%%%%%%%%%%%%%%%%%%%%%%%%
% BIBLIOGRAPHY
%%%%%%%%%%%%%%%%%%%%%%%%%%%%%%%%%%%%%%%%%%%%%%%%%%%%%%%%%%%%%%%%%%%%%%%%%%%

%\begin{thebibliography}{99}

%\end{thebibliography}

%%%%%%%%%%%%%%  FIGURES  %%%%%%%%%%%%%%%%%%%%%%%%%%%%%%%%%%%%%%%%%%%%%%%%%%

\begin{table}
\begin{tabular}{rrrr}
\multicolumn{1}{c}{Energy} & \multicolumn{1}{c}{Entropy LE} & 
\multicolumn{1}{c}{Entropy SD} & \multicolumn{1}{c}{$N_c$} \\
\cline{1-4} 
 0.00  &  -1.78525  &  -1.78247  & 79 \\
-0.12  &  -2.06341  &  -2.06124  & 56 \\
-0.20  &  -2.26269  &  -2.25417  & 44 \\      
-0.30  &  -2.50812  &  -2.50519  & 32 \\
\end{tabular}
\label{tab}
\end{table}

%%%%%%%%%%%%%%%%%%%%%%%%%%%%%%%%%%%%%%%%%%%%%%%%%%%%%%%%%%%%%%%%%%%%%%%%%%

\end{document}